\documentclass[12pt]{article}
\usepackage{graphicx}
\usepackage{bm}
\date{}
\begin{document}

\title{Combinatorial rules \\ of icosahedral capsid growth}

\maketitle

\author{Richard KERNER$^*$} \\

$*$  {Address : 
LPTMC, Universit\'e Paris-VI - CNRS UMR 7600 , \\ Tour 24, 4-\`eme , Boite 121,
4 Place Jussieu, 75005 Paris, France \\
Tel.: 33 1 44 27 72 98, \,  Fax: 33 1 44 27 51 00, \,  email : rk@ccr.jussieu.fr}

\vspace{0.6cm}
\indent
Abstract: \,  A model of growth of icosahedral viral capsids is proposed. 
It takes into account the diversity of hexamers' compositions, leading to definite 
capsid size. We show that the observed yield of capsid production implies a very high 
level of self-organization of elementary building blocks. The exact number of
different protein dimers composing hexamers is related to the size of a given
capsid, labeled by its $T$-number. Simple rules determining these numbers for
each value of $T$ are deduced and certain consequences are discussed.
\vspace{0.4cm}

\bigskip


\section{Introduction}
\indent
Just like the world of planets and stars, the world of viruses is ruled by
numbers. This is particularly true in the case of the numerous group of
{\it spherical viruses}, whose protective protein shells called ``capsids'' display  perfect
icosahedral symmetry \cite{Virology}. It is amazing that these structures, known to mathematicians 
since Coxeter's classification \cite{Coxeter}, are also observed in the 
so-called {\it fullerenes}, huge molecules composed exclusively of carbon 
atoms, predicted by Smalley and Kroto, and discovered in the eighties.  

Since Caspar and Klug \cite{CasparKlug} introduced simple rules predicting
a sequence of observed viral capsids, several models of growth dynamics 
of these structures have been proposed, e.g.  A. Zlotnick's model 
\cite{Zlotnick} published in 1994.  

The common geometrical feature of many viral capsids and fullerenes is their icosahedral
shape, with twelve pentagons found on the opposite sides of six five-fold symmetry axes,
and an appropriate number of hexagons in between. The number of hexagons is given by 
the following simple formula: $N_6 = 10 \, (T - 1),$ with $T = (p^2 + pq + q^2),$
called {\it the triangular number}, and where $p$ and $q$ are two non-negative integers 
\cite{Coxeter}. 

In capsids, the building blocks made of {\it coat proteins} are called {\it monomers}, 
{\it dimers}, {\it trimers}, {\it pentamers} and {\it hexamers}, according to their shape, 
the bigger ones usually being assembled from smaller ones prior to further agglomeration into 
capsid shells \cite{Stockley} . Sometimes pentameric or hexameric symmetry is displayed
despite the direct construction from $60$ or $180$ smaller subunits, like in the
{\it Cowpea mosaic virus} and the {\it Cowpea chlorotic mottle virus}, respectively
\cite{Larson}
Although certain virus species grow medium-size capsids corresponding to $N_6=20$ 
(like in the $C_{60}$ fullerene molecule), or $N_6=30$ and $N_6=60$, some of them form 
pure dodecahedral capsids (with exclusively pentamers as building blocks), like certain 
{\it Comoviridae} or {\it Cowpea} virus \cite{LinJohnson}), while some others, like human 
{\it Adenovirus} \cite{Stewart}, form very huge capsids with $N_6=240$, corresponding 
to $p=5, \, \, q=0$

In some cases, the similarity with the fullerene structure is striking: for example,
the TRSV capsid is composed of 60 copies of a single capsid protein (56 000 Da,
513 amino acid residues) \cite{Buckley}, which can be put in a one-to-one correspondence
with 60 carbon atoms forming a fullerene $C_{60}$ molecule; the aforementioned {\it Cowpea}
viruses provide another example of the same type. 

The process of building the icosahedral viral capsids differs quite essentially
from the fullerene formation: fullerenes are formed from carbon atoms and small carbon
molecules like $C_2$,$C_3$, up to $C_{9}$ or $C_{10}$), etc., in a hot plasma around  
electric arc between two graphine electrodes, whereas capsids are built progressively 
in liquid medium, from agglomerates of giant protein molecules displaying
pentagonal or hexagonal symmetry, or directly from smaller units ({\it monomers} or
{\it dimers}). It also seems that there is no such thing as universal assembly kinetics:
the way the capsids are assembled differs from one virus to another. The $T=7$ phage
HK47 appears to build pentamers and hexamers first, then assemble these capsomers to
form the final capsid structure, whereas another $T=7$ phage labeled P22 appears to
assemble its capsids directly from individual coat proteins (see \cite{Prevelige})
and the references within).

The common point is the presence of pentagons and hexagons in the resulting structure, 
and the strict topological rules that result from Euler's theorem on convex polyhedra:
$V - E + F = 2, \,$ with $V$ number of vertices, $E$ number of edges, and $F$  number 
of faces. From this one derives the fact that when only pentagonal and hexagonal faces are allowed, 
the number of pentagons is always $N_{5}=12$, while the hexagon number is $N_6=10(T-1)$.

Contrary to the case of fullerene molecules, whose yield from the hot plasma is in the best
case no higher than $10 \%$ of total mass of carbon sooth, viruses use
almost $100 \%$ of pentamers and hexamers at their disposal to form perfect icosahedral
capsid structures, into which their $DNA$ genetic material is densely packed
once the capsid is complete. 

This means that the initial nucleation ratio of pentamers versus hexamers is very close 
to its final value in capsids in order to minimize the waste. Secondly, the final size of 
the capsid must depend on particular assembly rules, which can be fairly well
deduced from the statistical weights of various agglomeration steps, found by maximizing
the final production rate. Let us investigate the rules that define the type and the
size of capsids, simultaneously optimizing the production rate.

\section{Probabilistic analysis of agglomeration }

The simplest stochastic model of growth successfully applied to fullerene formation
\cite{Kerner1} is based on the assumption that the dominant agglomeration processes
consist of forming new polygons in the cavities between two polygons on the
border of the existing cluster, by adjoining a $C_{2}$ or a $C_3$ molecule
to a cavity found in a cluster already formed. One of such
processes is shown in Fig.\ref{fig:Figure_2} below.

\begin{figure}[hbt]
\centering
  \includegraphics[width=5.2cm, height=4cm]{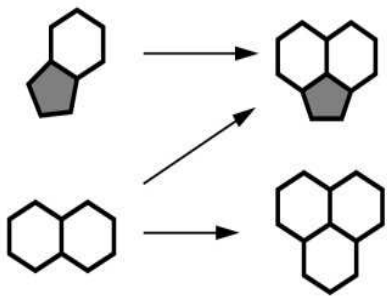}
\caption{\small{Creation of a new polygon in a cavity between two polygons}}
\label{fig:Figure_2}
\end{figure}

It is clear that the resulting $(666)$ cluster is wasted for further
fullerene formation, whereas a $(656)$ cluster can be used in the
next agglomeration step. Because the $(666)$ clusters are also absent
in final fullerene cage, it is easy to see that at each of consecutive
agglomeration steps the yield of "proper" clusters, useful for further 
fullerene construction, is exactly $1/2$. After about $23$ to $24$ steps leading
from the initial three-polygon structure to an almost finished fullerene cage
with 27 to 28 (out of total of 32) polygons already in place, the total
 yield would approach $2^{-24} \simeq 10^{-8}$ instead of observed $10^{-1}$,
i.e. $10\%$ ! This means that there is a mechanism that favours the creation
of ``correct'' structures versus the ``wrong'' ones, so that the average
yield of clusters proper for further fullerene construction becomes close to 
$q = 0.957$ at each agglomeration step, ensuring $q^{24}$ of order of $10^{-1}$.

In the case of fullerenes, the correction is due to the Boltzmann-Gibbs factors 
reflecting the energy differences between four basic processes : creating
a new pentagon in a $(6,6)$ cavity, or creating a new hexagon in a $(5,6)$
or in a $(6,6)$ cavity, assuming that the energy barrier against creation
of two or three pentagons sticking together is so big that the corresponding
Boltzmann factor is close to $0$. These factors could be evaluated by
requiring that the successive probabilities of finding pentagons among
all polygons in clusters of given size (after an n-th agglomeration step)
and the corresponding yields form a geometric progression \cite{Kerner1}

In the case of the icosahedral capsid formation the building process is not random
at all. One can be convinced that a high degree of self-organization is involved
by considering what would happen if even a small amount of randomness was
present. Let us exclude from our considerations the capsids formed exclusively
by pentamers; i.e. pure dodecahedral structures, and look at the build-up
of bigger capsids involving twelve pentamers and the necessary number of hexamers.

\begin{figure}[hbt]
\centering
  \includegraphics[width=6.4cm, height=5cm]{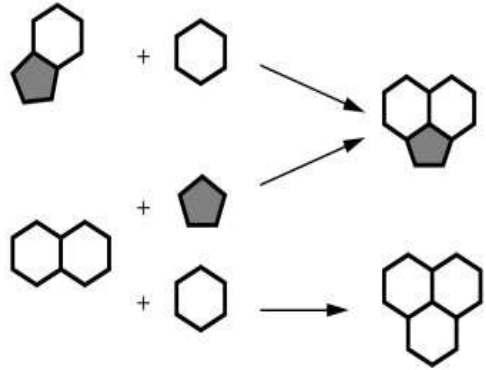}
\caption{\small{Adding a pentamer or a hexamer to an existing doublet}}
\label{fig:doublets}
\end{figure}

\begin{figure}[hbt]
\centering
  \includegraphics[width=6.4cm, height=7.2cm]{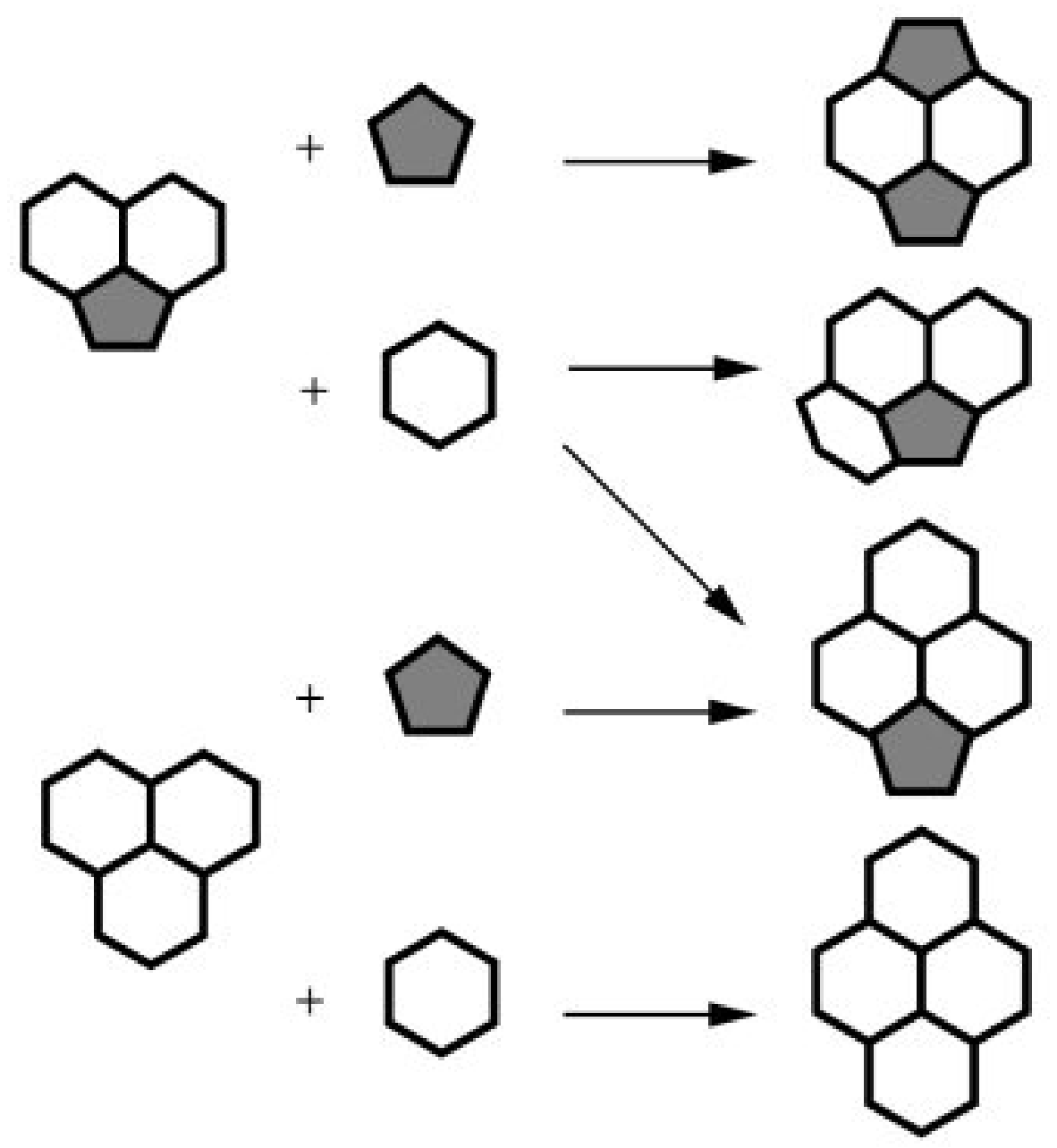}
\caption{\small{Next agglomeration step: from triplets to quadruplets}}
\label{fig:triplets}
\end{figure}
Let us denote the concentration (or the nucleation rate) of pentamers by $x$, that 
of hexamers by $(1-x)$. Then the probabilities of doublets are readily calculated as 
follows:
\begin{equation}
P_{56} = 2 \cdot W_{56} \, x(1-x)/ Q;  \ \ \, 
P_{66} =  W_{66} \, (1-x)^2 / Q,
\label{c1}
\end{equation}
where $W_{jk}$, $j,k = 5,6$ are the statistical weights depending on the 
virus type and on the chemical barriers between various sides, and 
$$Q = 2 \cdot W_{56} \, x(1-x) + W_{66}\, (1-x)^2$$ 
is the normalizing factor.
Note that we exclude two pentamers coming together, i.e. $W_{55} = 0$.
Similarly, the probabilities of three admissible "triplets" in the next 
agglomeration step displayed in Fig.\ref{fig:triplets} are given by:
\begin{equation}
P_{566} = P_{56} + 2 \cdot P_{66}\,  W_{66,5} \, x /Q_2, \, \ \ 
P_{666} = 2 \cdot P_{66}\,  W_{66,6}\,  (1-x)/Q_2,
\label{c2}
\end{equation}
where $W_{66,5}$ and $W_{66,6}$ denote the statistical weights of 
corresponding agglomeration processes, and $Q_2 = W_{66,5}\, x + W_{66,6} \, (1-x)$.
Now we can evaluate the average pentamer rate $x^{(k)}$ in clusters of given size, 
after $k$-th agglomeration step. The first three values are given by:
$$ x^{(1)} = \frac{1}{2} P_{56}, \, \ \ \, x^{(2)} = \frac{1}{3} P_{656}, \, \, \ \ 
x^{(3)} = \frac{1}{4} (2 P_{5656} + P_{6566} + P_{5666}),.. {\rm etc}$$
(In the expression for $x^{(3)}$ we use the probabilities for three different
allowed clusters, which are not discussed in detail here, but can be quite
easily obtained using the appropriate statistical weights and Boltzmann-Gibbs factors).

We can use these formulae in two different ways. Either we impose the statistical
weights $W_{ij}$ and $W_{ij,k}$, and determine the consecutive pentamer 
concentrations in growing clusters, starting from a given initial concentration $x ;$
or treat the statistical weights as unknowns and determine them from self-similarity
equations for successive pentamer concentrations:
\begin{equation}
\frac{x^{(n+2)} - x^{(n+1)}}{x^{(n+1)}-x^{(n)}} = \frac{x^{(n+1)} - x^{(n)}}{x^{(n)}-x^{(n-1)}} ,
\, \ \ \, n= 2,3,4...
\label{proportions}
\end{equation}
The resulting solutions for the limit values of $x$ and for auxiliary variables
$\xi = (W_{56})/(W_{66}), \, \eta =  (W_{56,6})/(W_{66,5}), \, \zeta = (W_{66,6})/(W_{66,5}),$ etc.,
although usually not in the form of simple fractions, give very good hints concerning
the assembly rules leading to particular capsid structures.

Now, as the capsid production rate from initial protein material is close to $100\%$,
for fullerene-like $(T=3)$ capsids we should have the initial rate of pentamers 
versus all capsomers as $3:8$. In order to keep the same ratio in the grown-up capsids, 
simple ``sticking rules'' will suffice. 

Let us denote pentamers' sides by symbol $p$, whereas two different kinds of sides
on hexamers' edges will be called $a$ and $b$ ( Fig. \ref{fig:BuildT3}). Suppose 
that a hexamer can stick to a pentamer with only $(p + a)$-combination; then
two hexamers must stick to each other only through a $(b + b)$ combination, with
both $(p + b)$ and $(a+b)$ combinations being forbidden by chemical potential
barrier. 
With these assumptions we get the following statistical factors: $ W_{56} = 15$, $W_{66} = 9$
$W_{56,6} = 3, W_{66,5} = 5$ and $W_{66,6} = 0$. With these rules the statistics 
in clusters will converge to the final value $x=3/8$ as shown in Fig.\ref{fig:BuildT3} 
below:

\begin{figure}[hbt]
\centering
  \includegraphics[width=4.3cm, height=4.3cm]{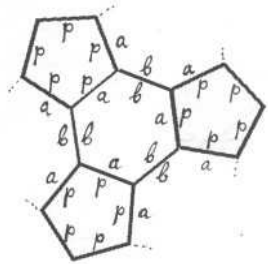}
\caption{\small{The building formula for the T=3 capsid}}
\label{fig:BuildT3}
\end{figure}

\begin{figure}[hbt]
\centering
  \includegraphics[width=4.6cm, height=4.6cm]{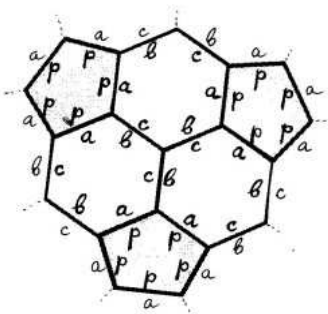}
\caption{\small{The building formula for the T=4 capsid}}
\label{fig:BuildT4}
\end{figure}

Similarly, with a more differentiate hexamer scheme,  $(abcabc)$, and with the assembling rules
allowing only associations of $p + a$ and $b+c$, we get with a $100\%$ probability
the $T=4$ capsid, with $x=2/7$, as shown in Fig. \ref{fig:BuildT4}. 

Note that in both cases we show only one of the ``basic triangles'' forming
the capsid, which is always made with $20$ identical triangles sticking
together to form a perfect icosahedral shape. 

 These examples suggest that strict association rules may exist providing
precise agglomeration pathways for each kind of icosahedral capsid. Let us analyze
these rules in more detail.

\section{Combinatorics of icosahedral capsids growth}

The virus capsid growth differs essentially from the fullerene agglomeration.
Taking into account the complexity of interactions between various proteins
forming hexamers and pentamers, the Boltzmann factors resulting from the energy
barriers should reduce to simple dichotomy: certain agglomerations are allowed,
whereas some others (like binding two pentamers together) are just forbidden. 
In contrast, the Boltzmann factors resulting from energy barriers between 
the allowed processes should not be very different at temperatures between
$20^o$ and $37^o$ C.

What remains then are pure statistical factors, which  must play the decisive role 
in order to ensure that the "correct" configurations are produced at each consecutive 
step almost without exception, i.e. practically with a $100 \%$ yield. Let us
show now how these statistical factors can be evaluated, and what constraints they
imply on the capsomer structure.

From symmetry considerations (and confirmed by chemical analysis) it results
that the pentamers are composed from five identical dimers, so that their five
edges are perfectly equivalent, and that they possess a defined orientation, i.e. 
it is known which one of the two faces will be on the outer side of the capsid.
All the five sides of a pentamer should be equivalent (identical), because $5$ is 
a prime number, and any division into parts will break the symmetry. 
Concerning the hexamers, as $6$ is divisible by $2$ and $3$,  one can have the following 
four situations:
\vskip 0.3cm
\indent
\hskip 0.5cm
- All $ 6$ sides equivalent, $(aaaaaa)$
\vskip 0.2cm
\indent
\hskip 0.5cm
    - Two types of sides, disposed  as  $(ababab)$
\vskip 0.2cm
\indent
\hskip 0.5cm
    - Three types of sides, disposed as  $(abcabc)$
\vskip 0.2cm
\indent
\hskip 0.5cm
    - Six different sides, $(abcdef)$
\vskip 0.2cm
\indent
The hexamers are also {\it oriented},  with one face becoming external,
and the other one turned to the interior of the capsid. The three differentiated
hexamers are represented in Figure \ref{fig:ababab} below.

\begin{figure}[hbt]
\centering
  \includegraphics[width=10cm, height=2.8cm]{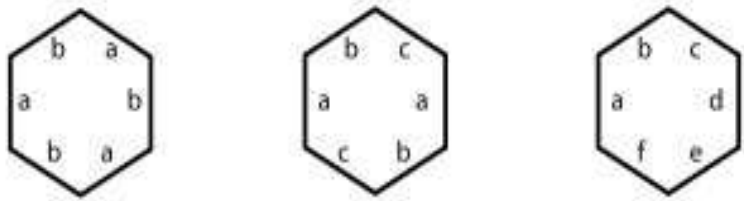}
\caption{\small{Three differentiations of hexamers}}
\label{fig:ababab}
\end{figure}

If the viruses were using undifferentiated hexamers with all their sides
equivalent, then there would be no reason for not creating any kind of
 structures as shown in Fig. \ref{fig:triplets} , and the final yield would be
very low (at best like in the fullerenes, less than $10 \%$). But with differentiated
hexamers of the $(ababab)-$ type simple selection rules excluding the $(p-b)$ and $(ab)$
associations while letting the creation of $(p-a)$ and of $(b-b)$ links, we have seen that 
the issue becomes determined with practically $100 \%$ certainty, as it follows from
the Fig. \ref{fig:BuildT3} .

These sticking rules can be summarized up in a table that we shall call the
{\it ``affinity matrix''} , displayed in the Table 1 below:

\begin{center}
\begin{tabular}{|c|c|c|c|}
\hline
\raisebox{0mm}[4mm][2mm]{ \, \ \ } & p & a & b \cr
\hline\hline
\raisebox{0mm}[4mm][2mm]{ p } & 0 & 1 & 0 \cr
\hline
\raisebox{0mm}[4mm][2mm]{ a } & 1 & 0 & 0 \cr
\hline
\raisebox{0mm}[4mm][2mm]{ b } & 0 & 0 & 1 \cr
\hline \hline
\end{tabular} 
\end{center}
\centerline{Table I : \, \small{Affinity matrix for the $T=3$ capsid construction}}
\vskip 0.2cm
Here a $``0''$ is put at the crossing of two symbols whose agglomeration is
forbidden, and a $``1''$ when the agglomeration is allowed. By construction,
a $``1''$ can occur only once any line or in any column. 
The next case presents itself when one uses the next hexamer type,
with a two-fold symmetry : $(abcabc)$. Again, supposing that only $a$-sides
can stick to pentamers' sides $p$, there is no other choice but the
one presented in Fig. \ref{fig:BuildT4}. The corresponding affinity matrix 
is as follows:

\begin{center}
\begin{tabular}{|c|c|c|c|c|}
\hline
\raisebox{0mm}[4mm][2mm]{ \, \ \ } & p & a & b & c \cr
\hline\hline
\raisebox{0mm}[4mm][2mm]{ p } & 0 & 1 & 0 & 0 \cr
\hline
\raisebox{0mm}[4mm][2mm]{ a } & 1 & 0 & 0 & 0 \cr
\hline
\raisebox{0mm}[4mm][2mm]{ b } & 0 & 0 & 0 & 1 \cr
\hline
\raisebox{0mm}[4mm][2mm]{ c } & 0 & 0 & 1 & 0 \cr
\hline \hline
\end{tabular} 
\end{center}
\centerline{Table II: \,\small{Affinity matrix for the $T=4$ capsid construction}}
\vskip 0.2cm

Finally, let us use the most highly differentiated hexamers of the $(abcdef)$-type.
Starting with pentamers surrounded by the hexamers sticking via the $(p-a)$-pairing,
we discover that now two choices are possible, leading to  left- and right-hand sided
versions, as shown in the following Fig. \ref{fig:BuildT7LR}

\begin{figure}[hbt]
\centering
\includegraphics[width=5.5cm, height=5.5cm]{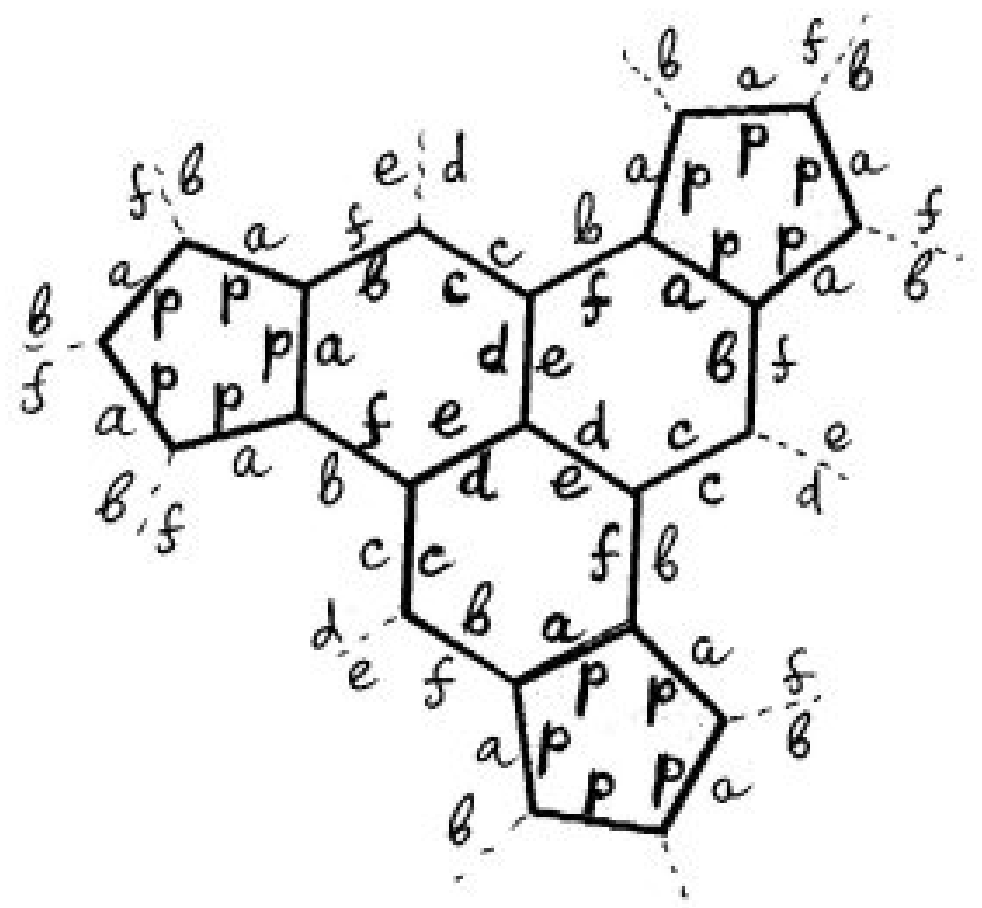}
\includegraphics[width=5.5cm, height=5.5cm]{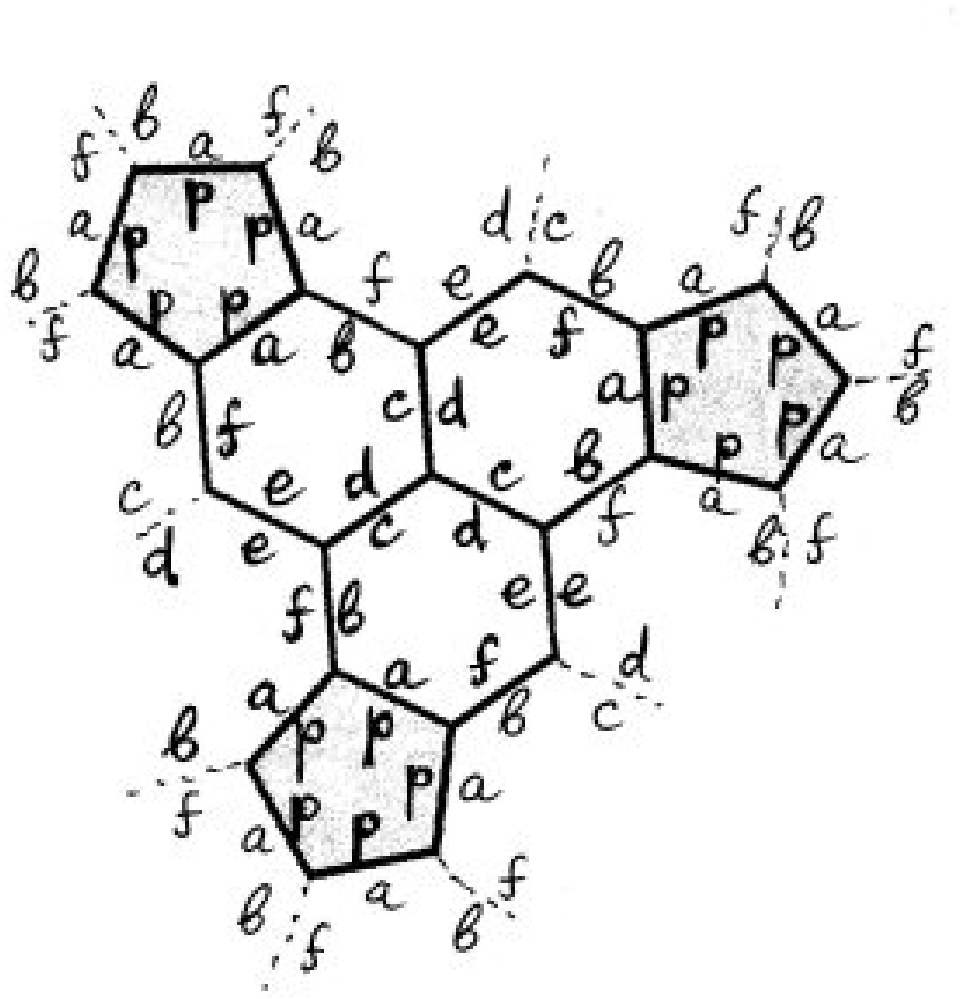}
\caption{\small{The building formulae for the T=7 capsids; left and right}}
\label{fig:BuildT7LR}
\end{figure}

The corresponding affinity matrices are given below:

\begin{center}
\begin{tabular}{|c|c|c|c|c|c|c|c|}
\hline
\raisebox{0mm}[4mm][2mm]{ \, \ \ } & p & a & b & c & d & e & f \cr
\hline\hline
\raisebox{0mm}[4mm][2mm]{ p } & 0 & 1 & 0 & 0 & 0 & 0 & 0 \cr
\hline
\raisebox{0mm}[4mm][2mm]{ a} & 1 & 0 & 0 & 0 & 0 & 0 & 0 \cr
\hline
\raisebox{0mm}[4mm][2mm]{ b } & 0 & 0 & 0 & 0 & 0 & 0 & 1 \cr
\hline
\raisebox{0mm}[4mm][2mm]{ c } & 0 & 0 & 0 & 1 & 0 & 0 & 0 \cr
\hline
\raisebox{0mm}[4mm][2mm]{ d } & 0 & 0 & 0 & 0 & 0 & 1 & 0 \cr
\hline
\raisebox{0mm}[4mm][2mm]{ e } & 0 & 0 & 0 & 0 & 1 & 0 & 0 \cr
\hline
\raisebox{0mm}[4mm][2mm]{ f } & 0 & 0 & 1 & 0 & 0 & 0 & 0 \cr
\hline \hline
\end{tabular} 
\end{center}
\centerline{Table III (a): \, \small{Affinity matrix for the $T=7$ (left) capsid construction}}
\vskip 0.2cm

\begin{center}
\begin{tabular}{|c|c|c|c|c|c|c|c|}
\hline
\raisebox{0mm}[4mm][2mm]{ \, \ \ } & p & a & b & c & d & e & f \cr
\hline\hline
\raisebox{0mm}[4mm][2mm]{ p } & 0 & 1 & 0 & 0 & 0 & 0 & 0 \cr
\hline
\raisebox{0mm}[4mm][2mm]{ a} & 1 & 0 & 0 & 0 & 0 & 0 & 0 \cr
\hline
\raisebox{0mm}[4mm][2mm]{ b } & 0 & 0 & 0 & 0 & 0 & 0 & 1 \cr
\hline
\raisebox{0mm}[4mm][2mm]{ c } & 0 & 0 & 0 & 0 & 1 & 0 & 0 \cr
\hline
\raisebox{0mm}[4mm][2mm]{ d } & 0 & 0 & 0 & 1 & 0 & 0 & 0 \cr
\hline
\raisebox{0mm}[4mm][2mm]{ e } & 0 & 0 & 0 & 0 & 0 & 1 & 0 \cr
\hline
\raisebox{0mm}[4mm][2mm]{ f } & 0 & 0 & 1 & 0 & 0 & 0 & 0 \cr
\hline \hline
\end{tabular} 
\end{center}
\centerline{Table III (b): \, \small{Affinity matrix for the $T=7$ (right) capsid construction}}
\vskip 0.2cm
Now a natural question can be asked: what comes next ? In order to grow capsids with $T$-numbers
greater than $7$, one has to introduce new types of hexamers that would never
stick to pentamers, but being able to associate themselves with certain sides
of the former maximally differentiated hexamers. The result is shown in the
Fig. \ref{fig:BuildT9} below:

\begin{figure}[hbt]
\centering
\includegraphics[width=4.7cm, height=4.7cm]{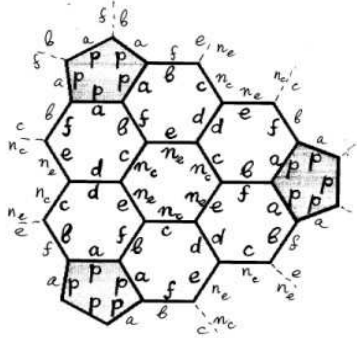}
\caption{\small{The building formula for the T=9 capsid}}
\label{fig:BuildT9}
\end{figure}

The corresponding affinity table is given below:

\begin{center}
\begin{tabular}{|c|c|c|c|c|c|c|c|c|c|}
\hline
\raisebox{0mm}[4mm][2mm]{ \, \ \ } & p & a & b & c & d & e & f & $n_a$ & $n_b$ \cr
\hline\hline
\raisebox{0mm}[4mm][2mm]{ p } & 0 & 1 & 0 & 0 & 0 & 0 & 0 & 0 & 0 \cr
\hline
\raisebox{0mm}[4mm][2mm]{ a } & 1 & 0 & 0 & 0 & 0 & 0 & 0 & 0 & 0 \cr
\hline
\raisebox{0mm}[4mm][2mm]{ b } & 0 & 0 & 0 & 0 & 0 & 0 & 1 & 0 & 0 \cr
\hline
\raisebox{0mm}[4mm][2mm]{ c } & 0 & 0 & 0 & 0 & 0 & 0 & 0 & 1 & 0 \cr
\hline
\raisebox{0mm}[4mm][2mm]{ d } & 0 & 0 & 0 & 0 & 1 & 0 & 0 & 0 & 0 \cr
\hline
\raisebox{0mm}[4mm][2mm]{ e } & 0 & 0 & 0 & 0 & 0 & 0 & 0 & 0 & 1 \cr
\hline
\raisebox{0mm}[4mm][2mm]{ f } & 0 & 0 & 1 & 0 & 0 & 0 & 0 & 0 & 0 \cr
\hline
\raisebox{0mm}[4mm][2mm]{ $n_a$ } & 0 & 0 & 0 & 1 & 0 & 0 & 0 & 0 & 0 \cr
\hline
\raisebox{0mm}[4mm][2mm]{ $n_b$ } & 0 & 0 & 0 & 0 & 0 & 1 & 0 & 0 & 0 \cr
\hline \hline
\end{tabular} 
\end{center}
\centerline{Table IV: \, \small{Affinity matrix for the $T=9$ capsid construction}}
\vskip 0.2cm

For bigger capsids, in which the rate of pentamers is lower, one can not obtain
proper probabilities unless more than one type of hexamers is present, out of which 
only one is allowed to agglomerate with pentamers. In the case of two different 
hexamer types one obtains either the $T=9$ capsid, or, with more exclusive
sticking rules, the $T=12$ capsid. Finally, in order to get the $T=25$ adenovirus
capsid, one must introduce no less than {\it four} hexamer types, out of which only
one type can agglomerate with pentamers.

\begin{figure}[hbt]
\centering
  \includegraphics[width=5.4cm, height=5.5cm]{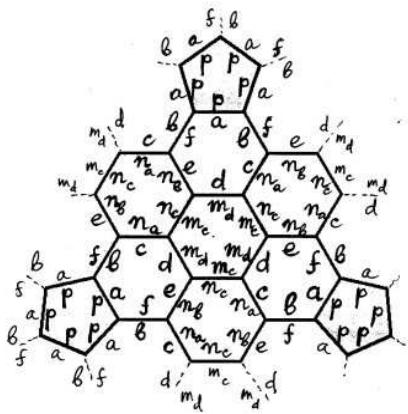}
\caption{The T=12 capsid's basic triangle}
\label{fig:BuildT12}
\end{figure}

The affinity matrix for $T=12$ capsid is as follows:
\begin{center}
\begin{tabular}{|c|c|c|c|c|c|c|c|c|c|c|c|c|}
\hline
\raisebox{0mm}[4mm][2mm]{ \, } & p & a & b & c & d & e & f & $n_a$ & $n_b$ & $n_c$
& $m_c$ & $m_d$ \cr
\hline\hline
\raisebox{0mm}[4mm][2mm]{ p } & 0 & 1 & 0 & 0 & 0 & 0 & 0 & 0 & 0 & 0 & 0 & 0 \cr
\hline
\raisebox{0mm}[4mm][2mm]{ a } & 1 & 0 & 0 & 0 & 0 & 0 & 0 & 0 & 0 & 0 & 0 & 0 \cr
\hline
\raisebox{0mm}[4mm][2mm]{ b } & 0 & 0 & 0 & 0 & 0 & 0 & 1 & 0 & 0 & 0 & 0 & 0 \cr
\hline
\raisebox{0mm}[4mm][2mm]{ c } & 0 & 0 & 0 & 0 & 0 & 0 & 0 & 1 & 0 & 0 & 0 & 0 \cr
\hline
\raisebox{0mm}[4mm][2mm]{ d } & 0 & 0 & 0 & 0 & 0 & 0 & 0 & 0 & 0 & 0 & 0 & 1 \cr
\hline
\raisebox{0mm}[4mm][2mm]{ e } & 0 & 0 & 0 & 0 & 0 & 0 & 0 & 0 & 1 & 0 & 0 & 0 \cr
\hline
\raisebox{0mm}[4mm][2mm]{ f } & 0 & 0 & 1 & 0 & 0 & 0 & 0 & 0 & 0 & 0 & 0 & 0 \cr
\hline
\raisebox{0mm}[4mm][2mm]{$n_a$} & 0 & 0 & 0 & 1 & 0 & 0 & 0 & 0 & 0 & 0 & 0 & 0 \cr
\hline
\raisebox{0mm}[4mm][2mm]{$n_b$} & 0 & 0 & 0 & 0 & 0 & 1 & 0 & 0 & 0 & 0 & 0 & 0 \cr
\hline
\raisebox{0mm}[4mm][2mm]{$n_c$} & 0 & 0 & 0 & 0 & 0 & 0 & 0 & 0 & 0 & 0 & 1 & 0 \cr
\hline
\raisebox{0mm}[4mm][2mm]{$m_c$} & 0 & 0 & 0 & 0 & 0 & 0 & 0 & 0 & 0 & 1 & 0 & 0 \cr
\hline
\raisebox{0mm}[4mm][2mm]{$m_d$} & 0 & 0 & 0 & 0 & 1 & 0 & 0 & 0 & 0 & 0 & 0 & 0 \cr
\hline \hline
\end{tabular} 
\end{center}
\centerline{Table V: \, \small{Affinity matrix for the $T=12$ capsid construction}}
\vskip 0.3cm
Now we can organize all these results in a single table that follows. To each value
of triangular number $T$ corresponds a unique partition into $1+(T-1)$, where the
$``1''$ represents the unique pentamer type and $(T-1)$ is partitioned into a sum of 
certain number of different hexamer types, according to the formula 
$$(T-1) = \alpha \, 6 + \beta \, 3 + \gamma \, 2$$
with non-negative integers $\alpha, \beta$ and $\gamma$. 

\begin{center}
\begin{tabular}{|c|c|c|c|}
\hline
\raisebox{0mm}[4mm][2mm]{Type (p,q)} & $T=p^2 + pq + q^2 $ & $N_6 = 10(T-1)$ &
$T$ decomposition  \cr
\hline\hline
\raisebox{0mm}[4mm][2mm]{(1,1)} & $3$ &  $20$ & $1+2$ \cr
\hline
\raisebox{0mm}[4mm][2mm]{(2,0)} & $4$ &  $30$ & $1+3$ \cr
\hline
\raisebox{0mm}[4mm][2mm]{(2,1)} & $7$ & $60$ & $1+6$ \cr
\hline
\raisebox{0mm}[4mm][2mm]{(3,0)} & $9$ & $80$ & $1+6+2$ \cr
\hline
\raisebox{0mm}[4mm][2mm]{(2,2)} & $12$ & $110$ & $1+6+2+3$ \cr
\hline
\raisebox{0mm}[4mm][2mm]{(3,1)} & $13$ & $120$ & $1+6+6$ \cr
\hline
\raisebox{0mm}[4mm][2mm]{(4,0)} & $16$ & $150$ & $1+6+6+3$ \cr
\hline
\raisebox{0mm}[4mm][2mm]{(3,2)} & $19$ & $180$ & $ 1+6+6+6 $ \cr
\hline
\raisebox{0mm}[4mm][2mm]{(4,1)} & $21$ & $200$ & $ 1+6+6+6+2 $ \cr
\hline
\raisebox{0mm}[4mm][2mm]{(5,0)} & $25$ & $240$ & $ 1+(4 \times 6) $ \cr
\hline 
\raisebox{0mm}[4mm][2mm]{(3,3)} & $27$ & $260$ & $ 1+ (4 \times 6)+2 $ \cr
\hline
\raisebox{0mm}[4mm][2mm]{(4,2)} & $28$ & $270$ & $ 1+ (4 \times 6)+3 $ \cr
\hline
\raisebox{0mm}[4mm][2mm]{(5,1)} & $31$ & $300$ & $ 1 + (5 \times 6) $ \cr
\hline 
\raisebox{0mm}[4mm][2mm]{(6,0)} & $36$ & $350$ & $ 1 + (5 \times 6) + 2 + 3 $ \cr
\hline 
\end{tabular} 
\end{center}
\centerline{Table VI: \, \small{Classification of icosahedral capsids. The last column gives}}
\vskip 0.1cm
\centerline{\small{ the number and type of hexamers needed for the construction}}

\newpage
\vskip 0.3cm
\indent
The inspection of  Table VI leads to the following simple rules:
\vskip 0.3cm
\indent
\hskip 0.5cm
 1) For the construction of a capsid with given triangular number $T$ one needs
exactly $T$ different proteins (or at least, $T$ different types of sticking sides) -
because the affinity matrix has the dimension $T \times T$, as easily seen in the
examples.
\vskip 0.2cm
\indent
\hskip 0.5cm   
2) By definition, the ``affinity matrices'' have only one non-zero item in each row
and in each column; moreover, they are symmetric. This means that {\it all} capsid
protein types (or more exactly, all different sticking sides) encountered in a complete 
icosahedron appear with the same frequency: $60$ times each. 
This can be most easily seen for the $p$-type forming a pentamer: there are
$12$ of them in each pentamer, and there are $12$ pentamers in any icosahedral capsid.
But then each $p$ sticks to an $a$, and exclusively to it: therefore, there must be
also $12 \, $ $a$-type proteins in the complete capsid, and so on, for each different protein.
This means that all the subunits that assemble in pentamers and hexamers later on have to be
produced at exactly the same rate in order to optimize capsid production.
\vskip 0.2cm
\indent
\hskip 0.5cm  
3) The capsomers composing a given capsid should be produced at different rates, with
a very simple rule: for every dozen of pentamers, one should have $60$ maximally diversified
hexamers of the type $(abcdef)$,  (and of each different type, like the $( n_a n_b n_c n_d n_e n_f )$
in the $T=9$ capsid); then $30$ hexamers of each $(ababab)$ type; and $20$ hexamers of the
$(abcabc)$ type. This rule can be easily seen upon inspection of Figures $4 - 8$.
 \vskip 0.2cm
\indent
\hskip 0.5cm   
4) In order to know how many (and of which kinds)  hexamers should be used,
 the $T$-number should be partitioned into $1 + $ the rest, the $"1"$ staying for the
unique type of pentamers' side, while ``the rest'' must be decomposed into
a sum of numbers $2$,  $3$ or $6$, according to the simple factors of $6$. This is shown
in the last column of Table I : we see that $T = 3 = 1 + 2; \, \ \ T = 4 = 1 + 3; \, \ \ $
then the next cases decomposes as $T=7 = 1 + 6; \,  \ \    T = 9 = 1 + 6 + 2$
(by the way, with this scheme in mind there is no point in trying to build a capsid
with $1 + 6 + 3 = 10 $ because $10$ cannot be found among the triangular numbers !).
\vskip 0.2cm
\indent
\hskip 0.5cm      
5) There is a clear "evolutionary" pattern in the last table - meaning that every next
(bigger)  type of capsid uses the previous construction, just adding a minimal amount
of novelty: and it is clear most of the time which kind of new hexamer one must add,
just looking at the differences between the consecutive T-numbers  - e.g. if they
differ by 2 or by 3, one should add one new hexamer type, ababab or abcabc,
respectively; but if they differ by 4 (e.g. from T=21 to T=25) or by 5 (from T=31 to T=36);
one must add  {\it two}  new types of $(ababab)$, or one $(ababab)$ and one $(abcabc)$ type.
\vskip 0.2cm
\indent
\hskip 0.5cm   
6) Finally, one should apply these reasonings also to capsids that are not built with
classical hexamers and pentamers. One may introduce a ``dual'' picture
in which not the sides, but the vertices of capsomers correspond to real proteins' extremities.
The examples of the alternative realisation of $T=3$ and $T=4$ capsids with pentamers and 
dimers only are shown in the Fig. \ref{fig:T34decorated}  below.

\begin{figure}[hbt]
\centering
\includegraphics[width=5.1cm, height=5.1cm]{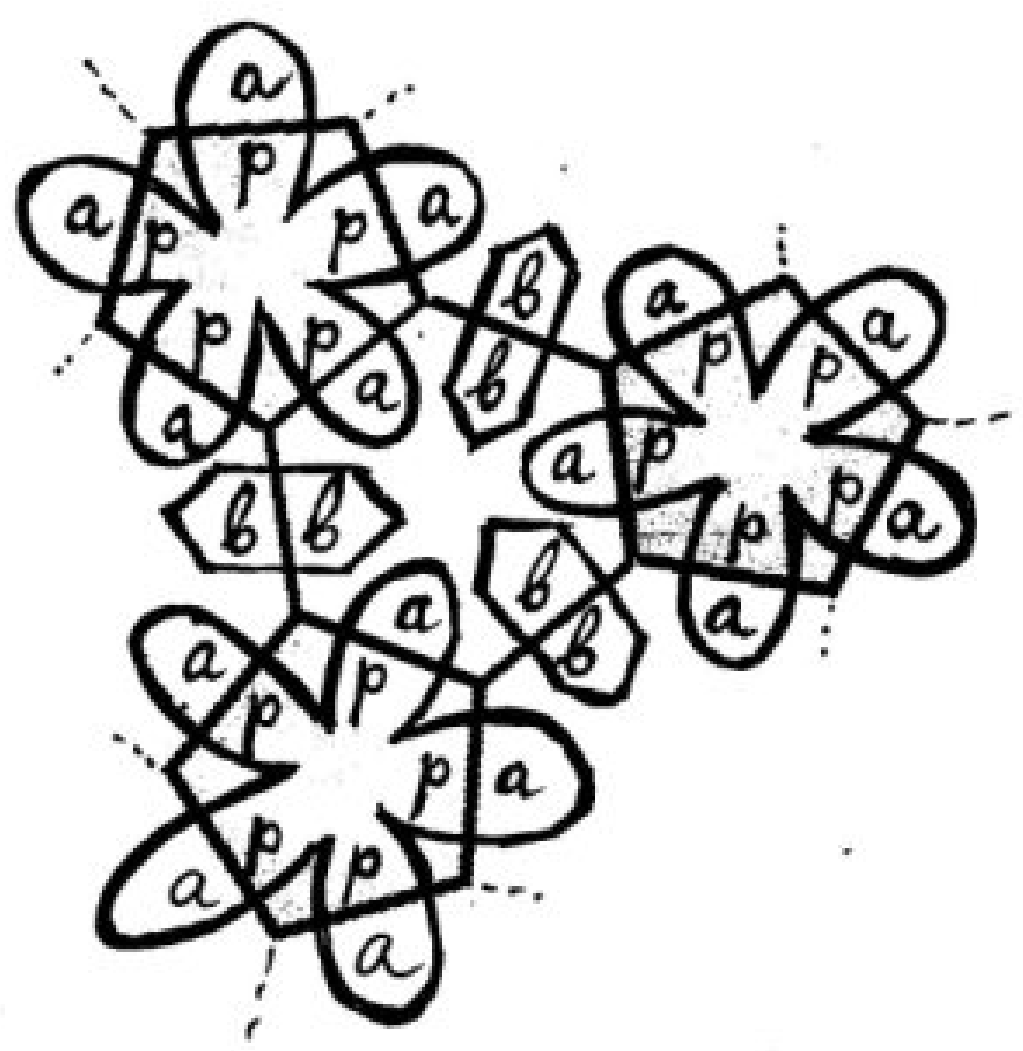}
\includegraphics[width=5.2cm, height=5.6cm]{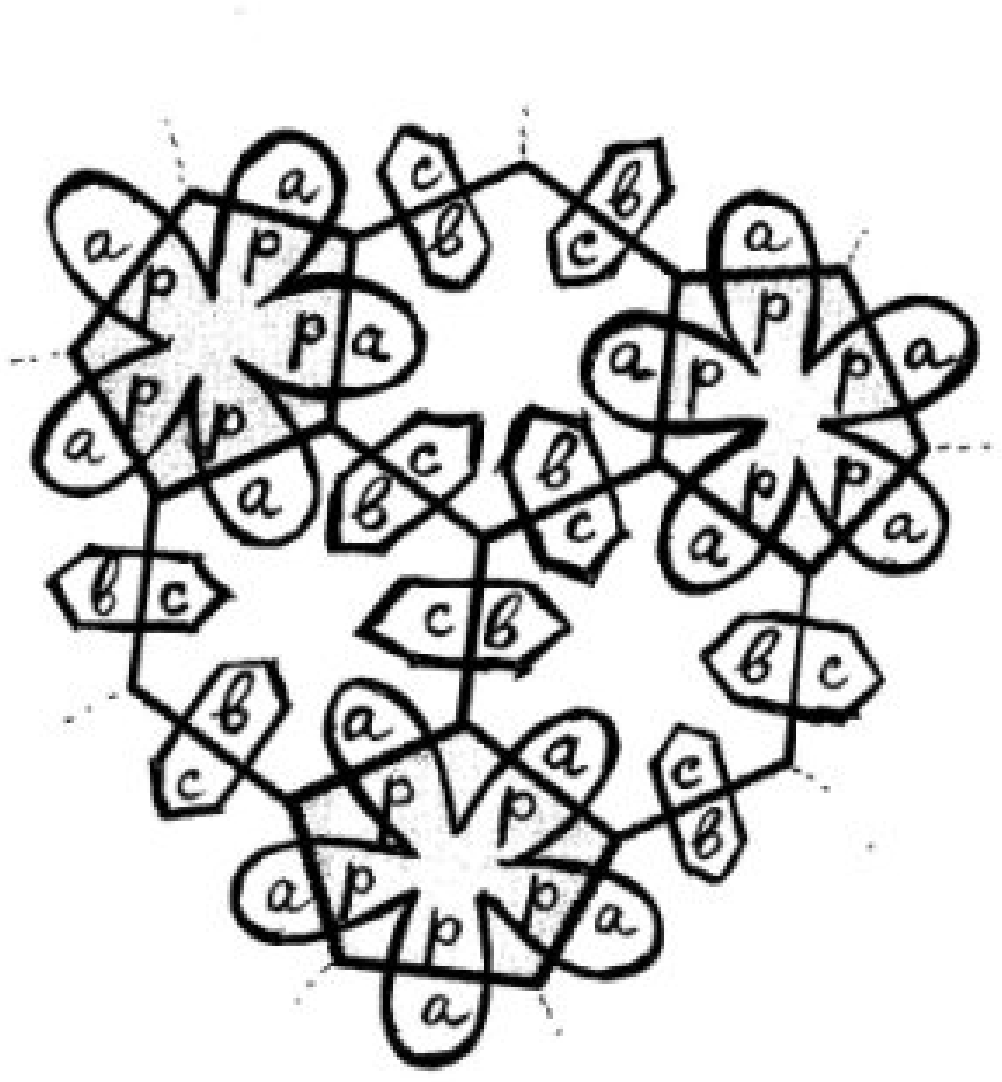}
\caption{\small{Alternative realizations of $T=3$ and $T=4$ capsids  
One can observe how the $2$- and $3$-fold symmetry axis do appear.}}
\label{fig:T34decorated}
\end{figure}

It can be easily deduced from these figures that the $(p-a)$-dimers composing 
pentamers occur $60$ times in each capsid, whereas the $(b-b)$-dimers occur
only $30$ times in a complete $T=3$ capsid, but the $(b-c)$-dimers occur
$60$ times in each $T=4$ capsid. In some cases in a $T=4$ capsid the triplets 
of $(b-c)$-dimers are replaced either by hexamers (then we obtain again 
a $T=3$ capsid), or by star-like {\it trimers} which will then occur $20$ times.
Whatever the decoration, each different letter symbol occurs with the same
frequency, i.e. $60$ times in each capsid independently of the value of $T$.
This suggests that all dimer proteins are produced at the same rate, and
the differentiation process that leads to exclusion rules for subsequent
agglomeration occurs later on. These realizations of capsid structure are akin 
to the decoration rules for curved Penrose-like tilings introduced by R. Twarock 
\cite{Twarock1}

\section{Conclusion: how to hinder capsid production ?}

Capsids are a vital part of viral life cycle, protecting its most essential
and the most vulnerable part, which is their genetic code. We need to consider
that viruses can have DNA or RNA genomes and the latter are especially sensitive
to degradation (by nucleases). In other words, without the capsid, 
the genomes exposed to the hostile external world would be destroyed in a couple
of hours by the ultraviolet radiation, or by chemical attacks of $SO_2$,
ozone $0_3$, $NO_2$, etc. The efficient protection provided by capsid shell
ensures longevity and makes possible long travels from one host to another.
Should we know how to hinder capsid assembly, tis information would have 
enormous potential benefit for the development of antiviral therapies.

The existence of a well-defined agglomeration scheme during the capsid
production suggests many possibilities of destructive intervention. As 
a matter of fact, the more complicated the system and the more intricate
the laws of its functioning, the more there are ways of hindering it.
Let us consider a few examples of how to decrease the efficiency of capsid
building.
\vskip 0.2cm
\indent
\hskip 0.5cm
1) Create seven-sided polygons - heptamers - with the same proteins that
usually form pentamers. Such a unit will be also surrounded by hexamers,
creating local {\it negative} curvature, thus enhancing the proper capsid
production. Such an achievement is quite unlikely becaus of the steric
hindrance - there is just not enough place in order to pack seven proteins
which usually go in packs by five. Even if there was such a possibility, 
the heptamers should be created inside the infected cells, or delivered there
in some way, which is extremely difficult.
\vskip 0.2cm
\indent
\hskip 0.5cm
2) A better way to hinder capsid production would be to produce hexamers or
pentamers with ``wrong'' proteins inserted, or in a wrong order - e.g. $(abbaab)$
or $(aaabbb)$ instead of $(ababab)$ so that they could fit with one
side, but then present a wrong protein to next capsomers trying to agglomerate,
thus destroying the symmetry and order of the construction. Again, this supposes
the creation of modified RNA chains ordering the production of different hexamers,
and again, the delivery problem seems very difficult to solve.
\vskip 0.2cm
\indent
\hskip 0.5cm 
3) A more natural way to hinder capsid production may be deduced from the probabilistic
analysis of various agglomeration pathways. Each capsid must contain exactly $12$
pentamers and $10 \, (T-1)$ hexamers of various kinds. In order to ensure the full
use of all these building blocks, their initial ratios should be as close as possible
to $12:(10(T-1))$. If for some reasons an excess of hexamers were produced,
entire capsids still would be completed leaving the extra hexamers unemployed.

   But the situation will radically change if an excess of pentamers could be created during
capsid assembly phase inside an infected cell. The agglomeration starts around the pentamers,
because the probability of a pentamer-hexamer association is much higher than that of 
a hexamer-hexamer association (\cite{Zlotnick}, \cite{Kerner1}). What will happen now can be 
illustrated on a concrete example. Let us imagine a great number of ``kits'' with 12 pentamers 
and 20 hexamers (ababab) each. As we know,  with simple matching rules allowing associations 
$(p+a)$ and $(b+b)$ and forbidding the associations  $(p + b)$ and $(a+a)$, a complete capsid 
can be constructed. Suppose now that many such ``kits'' have been dropped on the ground, and 
many people are trying simultaneously to build  $T=3$ capsids . 
The rules of the game being that once a person grabs a capsomer and sticks it to the partially 
built capsid, the capsomer can not be removed. After some time everybody will succeed in
constructing an entire full capsid, with no extra capsomers left.
    
    Suppose now that someone had thrown in some extra hexamers. After a while some of the hexamers 
will be left out - but if there were $12 N$ pentamers and more than $20 N$ of hexamers, there will 
be still $N$ full capsids completed at the end of the day.
    
   The situation would be totally different had someone poured in an excess of pentamers. 
Supposing that people grab capsomers one by one at random, there will be
 hardly one or two full capsids completed.
    
This comes out from simple probability calculus. Let us start
 with the simplest example, $20$ hexamers $(ababab)$ and 13 pentamers
$(ppppp).$ Of course, if the things were not happening at random, but in an
organized way, one can construct a complete $T=3$ capsid and leave the extra
 pentamer alone. There are exactly $13$ ways of doing it - just taking the decision
 which one among the $13$ pentamers should be chosen to be dropped out.
    But if the agglomeration happens at random, there is no reason that
 one of the pentamers should wait until the $12$ others are incorporated in
 a capsid. People acting randomly would rather pick up as many hexamers as they can; and
 there is an enormous number of ways of doing it (with all possible
permutations taken into account), so the probability of this to happen is orders 
of magnitude higher than the probability of the happy issue described above. 
For example, there are $13 \times 20$ different ways of producing two incomplete
capsids: a pentamer-hexamer doublet and an incomplete $T=3$ capsid with one
hexamer missing; and there are $13 \times 20 \times 19$ different ways of
producing a $5-6-6$ agglomerate and an incomplete capsid with two hexamers
missing, and so on.

This means that as a result, we shall have in the best case {\it two} uncomplete capsids, 
and most probably even a higher number of incomplete, unfinished structures. 
  
This is what would happen to people trying to pick up spare parts
at a car cemetery, full of naked car frames and spare wheels, and to make up a complete car. 
``Complete'' meaning four wheels in one car - with three wheels only no car will ever
roll out. But if the number of wheels is not $4$ times $N$ ($N$ being the
number of car frames with no wheels), but only half of it, say, then what 
would most probably happen is that almost everybody would
 end up with a car with 3 or 2 wheels, and almost nobody with a complete car.
   
 In other words, if we could incite some cells to produce
{\it exclusively  pentamers}  of the virus by which it is attacked,  - this has to
happen inside the cell, where the capsids are being produced - then there
 will be an excess of pentamers, and almost no complete capsid will be
 produced. And with no complete capsids viruses will be much less
obnoxious.

  To implement such a scenario is certainly a challenge for molecular biology
and nano-technology.

\section*{Acknowledgments}
The author is greatly indebted to Dr. Nicola Stonehouse for many enlightening discussions
and for her extremely useful suggestions and remarks. Thanks are also due to  Drs. Reidun Twarock 
and Adam Zlotnick for interesting discussions, and to Dr. Maja Nowakowski for her help in 
getting acquainted with current literature and careful reading of the manuscript.

\newpage

\end{document}